\documentclass[prd,showpacs,preprintnumbers,amsmath,amssymb]{revtex4}
\usepackage{amsmath,amssymb,graphicx} 
\usepackage {amssymb}
\newcommand{\nc}{\newcommand}
\nc{\ba}{\begin{eqnarray}}
\nc{\ea}{\end{eqnarray}}
\newcommand\be{\begin{equation}}
\newcommand\ee{\end{equation}}
\newcommand\bb{{\mathbf{b}}}
\newcommand\baa{{\mathbf{a}}}

\newcommand\bx{{\mathbf{x}}} 
\newcommand\si{\sigma}
\newcommand\al{\alpha}

\nc{\ga}{\gamma}
\nc{\tnu}{\bar{\nu}}
\nc{\tmu}{\bar{\mu}}
\nc{\tq}{\tilde{q}}

\newcommand{\la}{\lambda}
\nc{\x}{{\bf{x}}}
\nc{\bmi}{\bar \mu_{i} }
\nc{\bm}{\bar \mu }
\newcommand{\gsim}{\raise.3ex\hbox{$>$\kern-.75em\lower1ex\hbox{$\sim$}}}
\newcommand{\lsim}{\raise.3ex\hbox{$<$\kern-.75em\lower1ex\hbox{$\sim$}}}
\nc{\gm}{\gamma }

\input{epsf.sty}

\begin{document}

\rightline{Imperial/TP/07/TK/02}
\title{On the Collision of Cosmic Superstrings}
\author{E.~J.~Copeland}
\email{ed.copeland@nottingham.ac.uk} \affiliation{School of
Physics and Astronomy, University of Nottingham, University Park,
Nottingham NG7 2RD, United Kingdom}

\author{H.~Firouzjahi}
\email{firouz@hep.physics.mcgill.ca} 
\affiliation{
Physics Department, 
McGill University, 
3600 University Street,
Montreal, H3A 2T8, Canada.}

\author{T.~W.~B.~Kibble}
\email{kibble@imperial.ac.uk}
\affiliation{Blackett Laboratory, Imperial College, London SW7 2AZ, United 
Kingdom}

\author{D.~A.~Steer}
\email{steer@apc.univ-paris7.fr} \affiliation{APC, University Paris 7,
10, Rue Alice Domon et L\'eonie Duquet, 75205 Paris Cedex 13,
France}

\begin{abstract}
We study the formation of three-string junctions between $(p,q)$-cosmic superstrings, and collisions between such strings and show that kinematic constraints analogous to those found previously for collisions of Nambu-Goto strings apply here too, with suitable modifications to take account of the additional requirements of flux conservation.  We examine in detail several examples involving collisions between strings with low values of $p$ and $q$, and also examine the rates of growth or shrinkage of strings at a junction.  Finally, we briefly discuss the formation of junctions for strings in a warped space, specifically with a Klebanov-Strassler throat, and show that similar constraints still apply with changes to the parameters taking account of the warping and the background flux.

\end{abstract}

\maketitle
\section{Introduction}

One of the most plausible scenarios leading to the formation of cosmic strings  
\cite{Sarangi:2002yt, Jones:2003da}
is that of brane inflation \cite{Dvali:1998pa}. 
In these models, the inflaton field is the distance between a D3-brane and an anti-D3-brane.  There is an attractive Coulombic potential between them, which may lead to 
a slow-roll inflation. In models of warped brane inflation \cite{Kachru:2003sx}, the fine-tunning problem associated with the flatness of the potential in the original models of brane inflation may be less severe. Furthermore,  due to warping the tension spectrum of cosmic superstrings, $\mu$, is much smaller than the naive expectation, $m_{s}^{2}$, where $m_{s}$ is the mass scale of string theory. Consequently, $G \mu$, the dimensionless number measuring the tension of cosmic superstrings in units of the Newton constant, $G$, can be lowered below the current observational bounds 
\cite{Firouzjahi:2005dh, Wyman:2005tu}.

A brane collision creates both fundamental (F-) strings and D-strings (D1-branes), as well as $(p,q)$-strings, composites of $p$ F-strings and $q$ D-strings \cite{Copeland:2003bj, Jackson:2004zg}. 
When cosmic superstrings of different types collide, they can not intercommute, instead they exchange partners and form a junction. This is a simple consequence of charge conservation at the junction of 
colliding $(p,q)$-strings and is in contrast to ordinary Abelian gauge strings, where two colliding strings
usually exchange partners and intercommute with the intercommutation probability of order unity. 
Most work done to date on cosmic strings has involved such abelian strings because of the possibility that they could have formed in the early universe as a result of phase transitions possibly associated with the GUT scale. The interest in strings with junctions has been more at the level of a curiosity although earlier work has considered them in relation to monopoles connected to strings \cite{Vachaspati:1986cc} and in terms of non-abelian gauge theories \cite{Spergel:1996ai,McG}. However, it has been the recent realisation that there could be superstrings of comic length that has prompted people to take seriously the implications of junction forming strings, in particular could there be observational smoking guns that would enable us to determine whether they really exist? It is primarily  this prospect which motivates us to consider the kinematics associated with $(p,q)$ strings. 

Previous studies \cite{Copeland:2006eh,Copeland:2006if, Brandenberger:2007ae} dealt with three-string junctions between Nambu-Goto (NG) strings. However $(p,q)$-strings carry fluxes of a gauge field and are therefore described instead by the Dirac-Born-Infeld (DBI) action.  Constraints arise not only from conservation of energy and momentum but also from charge conservation.

Our aim here is to study how this difference affects the behaviour of strings at junctions.  A very important feature of the DBI action is that, as in the NG case, it is reparametrization invariant; because of that, many of the previous results carry over, but with important changes.  Throughout, we work in the probe-brane approximation, neglecting back-reaction on the space-time metric. The layout of the paper is as follows: in section \ref{junctions} we derive the required equations of motion and charge conservation conditions for $(p,q)$ strings at a junction; and in section \ref{temporal} we show that it is consistent to write everything in the temporal gauge. This is followed in section \ref{constraints} by a detailed analysis of the collision of strings with different charges, meeting at an arbitrary angle and relative velocity. In particular we derive analytic expressions which show the conditions under which junctions can form in terms of the string tensions, as well as their angle and speed of approach, including some specific examples. In section \ref{warped} we extend our  analysis to include the collision of cosmic superstrings in a warped throat, making the calculation applicable to the string motivated models of warped brane inflation. Finally we conclude in section \ref{conc}, where some comments on average properties of the network are also given.

\section{Equations of motion and charge conservation \label{junctions}}

In this section we set up the equations of motion for charged $(p,q)$-strings meeting at a junction, and discuss the question of gauge fixing.  Consider first a single $(p,q)$ string evolving in a flat target space-time metric $\eta_{\mu \nu}$ ($\mu=0,\dots,3$), with position $X^\mu(\sigma^a)$ where the two worldsheet coordinates are $\sigma^a=(\tau,\sigma)$.  The induced metric on the string worldsheet, $\gamma_{ab}$, is given by
\ba
\gamma_{ab}= \eta_{\mu \nu} X^{\mu}_{,a}  X^{\nu}_{,b} .
\ea
Similarly to the Nambu-Goto action, the DBI action enjoys world-sheet reparametrization invariance.  Thus two of the three degrees of freedom of $\gamma_{ab}$ can be 
eliminated by imposing the conformal gauge; that is choosing $\{\tau, \sigma\}$ such that $\gamma_{ab}$ exhibits explicitly the conformal flatness common to all 2D spaces:
\ba
\label{conformal}
\dot X^{2} + X'^{2}=0 , \qquad  \dot X \cdot  X'=0.
\ea
The prime and over-dot denote derivatives with respect to $\sigma$ and $\tau$ respectively and the inner products in (\ref{conformal}) is defined via the metric $\eta_{\mu \nu}$ with the signature
$(+,-,-,-)$.  
 As we shall show later, it is also consistent to impose the temporal gauge condition, choosing the temporal world-sheet coordinate $\tau$ to coincide with the global time coordinate $X^0$.  For the moment we do not impose this condition.

As discussed in \cite{Witten:1995im,Polchinski}, a $(p,q)$-string is a bound state of $p$ fundamental strings, F-strings, and $q$ D1-brane, D-strings, where $p$ and $q$ are integer coprime numbers.  Alternatively
one can view the $(p,q)$-string with $q\neq 0$ as $p$ units of electric flux dissolved on the world volume of $q$ coincident D-strings.  In flat-space time and in the absence of the background fluxes the Chern-Simons terms  vanish and the action is given by the DBI part. Working in the conformal gauge (\ref{conformal}), the action for the $(p,q)$-string is \cite{Polchinski}
\ba
\label{action1}
S= - \mu   \int  d \tau  \, d \sigma \, \sqrt{- | \gamma_{ab}+ \lambda F_{ab} | }=
- \mu   \int  d \tau  \, d \sigma \, \sqrt{ - X'^2  \dot X^{2}  
- \la^2 F_{\tau\sigma} ^2}    ,
\ea
where $\mu={|q|}/{ (g_{s} \la)}$ is the tension of  $q$ coincident D-strings,  $\la= 2 \pi \alpha'$ and
$g_{s}$ is the perturbative string coupling.  The electric flux $p$ is given by $p=\partial {\cal L}/ \partial F_{\tau \sigma}$ where $S=   \int   d \sigma \, {\cal L}.$

Now consider three cosmic $(p_i,q_i)$-strings (labeled by index $i=1,2,3$) meeting at a junction.  The action for the system is 
\ba
\label{action}
S&=& -\sum_{i} \mu_i   \int  d \tau  \, d \sigma \, \sqrt{ - X_i'^2  \dot X_{i}^{2}  
- \la^2 { F_{\tau\sigma}^i} ^2}    
\theta \left(s_{i}(\tau)-\sigma \right)  \nonumber\\
&+& \sum_{i}  \int d  \tau  \left(    f_i \cdot \left[  X_i(s_i (\tau),  \tau )-\bar X(\tau)  \right]  +
g_i  \left[  A^i_\tau (s_i(\tau),\tau)+\dot s_i   A^i_\sigma(s_i(\tau),\tau)
 - \bar A(\tau) \right] \right)
\ea
where $f^\mu_{i}$ and $g_{i}$ are the Lagrange multipliers, and
\be
\mu_{i}= \frac{|q_{i}|}{ g_{s} \la}.  
\ee
The $f^\mu_i$ term imposes that the position $X^\mu_i(s_i(\tau),\tau)$ of the end-point of each of the three strings (as imposed by the theta function in $\sigma$) coincides with the position ${\bar X}^\mu(\tau)$ of the vertex.  The $g_i$ term performs a similar task for the U(1) gauge potential $A^i_a$, ensuring that its component tangent to the vertex worldline coincides with the single component of the gauge potential $\bar A$ confined to the worldline.
The action is invariant under the U(1) gauge transformation 
$A^{i}_{a} \rightarrow A^{i}_{a} - \partial_{a}  \Lambda^{i}$ for $a= (\tau, \sigma)$, 
provided that we also make the transformation 
 $$
 \bar A\to\bar A-(d/d\tau)\bar\Lambda,
 $$
with the identification
$$ \bar\Lambda(\tau)=\Lambda^i(s_i(\tau),\tau), $$
which ensures that the $\Lambda^i$ from the different legs match up correctly.

The equations of motion for the gauge field components $A^{i}_{\tau}$ and $A^{i}_{\sigma}$
 respectively are (no summation over $i$)
\ba
\label{A0eom}
\mu_{i }   \partial_{\sigma}   \frac{  \la^{2}   F^i_{\tau \sigma}    \theta \left(s_{i}(\tau)-\sigma \right)   }{ \sqrt{ - X_i'^2  \dot X_{i}^{2}  - \la^2 { F_{\tau\sigma}^i} ^2}   }= -g_{i}   \delta( s_{i}- \sigma),
\ea
\ba
\label{Aseom}
\mu_{i }   \partial_{\tau}   \frac{ \la^{2}    F^i_{\tau \sigma}    \theta \left(s_{i}(\tau)-\sigma \right)   }{ \sqrt{ - X_i'^2  \dot X_{i}^{2}  - \la^2 { F_{\tau\sigma}^i} ^2} }= g_{i}   \dot s_{i}   \delta( s_{i}- \sigma).
\ea
The equation of motion for $X^{\mu}$ is 
\ba 
\mu_{i}  \left(\partial_{\tau} \frac{X_i'^{2} \dot X_{i}^{\mu}   \theta \left(s_{i}(\tau)-\sigma \right)  }
{ \sqrt{ - X_i'^2  \dot X_{i}^{2}  - \la^2 { F_{\tau\sigma}^i} ^2}  }  + 
\partial_{\sigma} \frac{\dot X_i^{2}  X_i'^{\mu}   \theta \left(s_{i}(\tau)-\sigma \right)  }
{ \sqrt{ - X_i'^2  \dot X_{i}^{2}  - \la^2 { F_{\tau\sigma}^i} ^2}  }  \right)
= f_{i} ^{\mu}\delta( s_{i}- \sigma).
\ea
The electric flux $p_{i}$ along each string, determining the number of F-strings on the D-string world volume,  is 
given by $p_i =\delta {\cal L}/ \delta F_{i\, \tau \sigma}$, which gives
\ba
\label{F}
p_{i}=  \frac{ \la^{2}   \mu_{i}    F^i_{\tau \sigma}       }{ \sqrt{ - X_i'^2  \dot X_{i}^{2}  - \la^2 { F_{\tau\sigma}^i} ^2}  }  .
\ea
From equations  (\ref{A0eom}) and (\ref{Aseom})  we see 
that $\partial _{\tau} p_{i}= \partial _{\sigma}p_{i}=0$.  This is a manifestation of the fact that in 2-D the
gauge field equations are trivial.

Now define
\ba
\label{barmu}
\bar \mu_{i} = \sqrt{ \lambda^{2}   \mu_{i}^{2} +p_{i}^{2}}  = \sqrt{ p_{i} ^{2}  + \frac{q_i^2}{g_{s}^2 }  }
\ea
which, as we show below, is nothing other than the tension of the $i^{\rm th}$ string (up to a factor of 
$\lambda$).
Then from Eq.~(\ref{F})  one obtains that
\ba
\label{F2}
\la  F^i_{\tau\sigma}= -\frac{p_{i} }{ \bmi}     {X_{i}' }^{2} ,
\ea
where the minus sign comes from the fact that in Eq.~(\ref{F}) both $F^{i}_{\tau \sigma}$ 
and $p_{i}$ have the same sign while from Eq.~(\ref{conformal}) we have $ {X_{i}' }^{2} = -\dot X_{i}^{2} <0 $.
Furthermore
\ba
\label{sqrt}
 \sqrt{-  X_i'^2  \dot X_{i}^{2}  - \la^2 { F_{\tau\sigma}^i} ^2} = - \frac{ \la \mu_{i} }{\bmi  }   {X_{i}' }^{2}  .
\ea
Using this in the equation of motion for $X^{\mu}$ we find
\ba
\label{Xeom}
\ddot{X}_i^{\mu} -{X_{i}^{\mu} }''=0  ,
\ea
with the following boundary condition at $\sigma= s_{i}(\tau)$:
\ba
\label{f1}
\bmi   ({ X_{i}^{\mu} }' + \dot s_{i} \dot X_{i}^{\mu}) = - \la f_{i}^{\mu}  .
\ea
Similarly, the boundary conditions from Eqs.~(\ref{A0eom}) and (\ref{Aseom}) imply
\ba
\label{Abc}
g_{i} = p_{i}   .
\ea

On the other hand, varying the action with respect to $\bar  X, \bar A_{\tau}$ and $\bar A_{\sigma}$, respectively, yields
\ba
\label{f2}
\sum_{i} f_{i}^{\mu}= \sum_{i} g_{i} =0  .
\ea
Using these in equations  (\ref{Abc}) and (\ref{f1}), one obtains the following conservation laws:
\ba
\label{fcons}
\sum_{i} \bmi   ({ X_{i}^{\mu} }' + \dot s_{i} \dot X_{i}^{\mu})=0  ,
\ea
and
\ba
\label{pcons}
\sum_{i} p_{i}=0    .
\ea
Physically (\ref{pcons}) indicates the electric flux conservation at the junction. With this convention, all fluxes are flowing {\it into} to the junction.

Varying the action with respect to the Lagrange multipliers $f_{i}^{\mu}$ and  $g_{i}$ respectively imposes the desired constraints for the fields at the junction  $\sigma= s_{i}(\tau)$;
\ba
\label{Lagrange}
X_{i}^{\mu}(s_{i}(\tau), \tau)= \bar X^{\mu}, \qquad 
A_{i \tau} (s_{i}(\tau), \tau) + \dot s_{i}  A_{i \sigma} (s_{i}(\tau), \tau)= \bar A(\tau),
\ea
where the second equation is subject to the U(1) gauge invariance as described previously.
Finally, one can check that the equation coming from the variation of the action with respect to $s_{i}$ is automatically satisfied
\ba
\frac{\delta S}{\delta s_{i}} = \int d   \tau \left( - \mu_{i}    
\sqrt{ - X_i'^2  \dot X_{i}^{2}  - \la^2 { F_{\tau\sigma}^i} ^2} + 
f_{i  \mu} X_{i}'^{\mu} + g_{i}  A_{i \tau} ' - g_{i}  \partial_{\tau} A_{\sigma} \right)  = 0   ,
\ea
where all quantities labeled by indices $i$ are evaluated at $\sigma = s_{i}(\tau)$. To obtain
the final result equations (\ref{F2}),  (\ref{sqrt}), (\ref{f1}) and (\ref{Abc}) 
were used. 

Motivated by electric flux conservation (\ref{pcons}), one expects the same conservation law for the D-string charges:
$\sum_{i} q_{i} =0$. 
To see this, recall that D-strings are charged under the Ramond-Ramond 
two-form $C_{(2)}$ via the Chern-Simons part of the action \cite{Polchinski}
\ba
\label{CS}
S^i_{CS}= \frac{q_i}{2 \lambda } \int d   \tau \, d \sigma  \, \epsilon^{a b}   \partial_{a} 
X^{\mu}_i    \partial_{b} X^{\nu}_i C_{(2)  \mu \nu}.
\ea
%
The  action also contains the kinetic energy for $C_{(2)}$. This action is invariant
under the gauge transformation 
\ba
\label{Cgauge}
C_{(2) \mu \nu} \rightarrow C_{(2)  \mu \nu} + \partial_{\mu} \Lambda_{\nu} -  
\partial_{\nu} \Lambda_{\mu} .
\ea
Under this gauge transformation, the sum of the Chern-Simons terms for the three strings at the junction transforms according to
$$
\delta S_{CS}=  -\sum_{i} \frac{q_{i}}{\lambda }  
\int d   \tau  \Lambda_{\mu} \frac{d X_{i}^{\mu}(s_i, \tau)}{ d\tau} 
= -\sum_{i} \frac{q_{i}}{\lambda }    \int d   \tau  \Lambda_{\mu} \frac{d \bar X^{\mu} }{ d\tau}.
$$
In order to keep the action invariant under the gauge transformation (\ref{Cgauge}), we require
\ba
\label{qcons}
\sum_{i} q_{i} =0,
\ea
as expected.

In our analysis so far we assumed that $q_{i}\neq 0$ so a $(p,q)$-string is obtained by turning $p$ units
of electric flux on the world volume of $q$ D-strings. If $q=0$ for some strings, then in Eq. (\ref{action1}) one replaces $\mu$ by the tension of $p$ F-strings, $p/\lambda$, with no gauge field on the world volume.
All our results, namely equations (\ref{pcons}) and (\ref{qcons}) go through, while for $(p,0)$ strings, 
$\bm$ is still given by (\ref{barmu}) with $q=0$.


\section{\bf Temporal gauge\label{temporal}}

So far we have imposed the conformal gauge (\ref{conformal}): one may ask whether this fixes the gauge completely, or whether (as for NG strings in Minkowski space) it is also possible to impose the temporal gauge condition $X^{0}=t=\tau$.  In this section we show that the temporal gauge is consistent with the above equations of motion.

In the temporal gauge conditions (\ref{conformal}) reduce to 
\ba
\label{temp}
 {{\dot \x_{i}}^{2} } + {\x_{i}'^{2} }=1 , \qquad { {\dot \x_{i}} \cdot \x_{i}' }=0  ,
\ea
where 
$X_{i}^{\mu}=(t,\x_{i})$.
Consider first the $\mu=0$ component  of (\ref{fcons}) which, in the temporal gauge (with $\cdot=d/dt$) gives the constraint
\ba
\label{energy}
\sum_{i} \bmi   \dot s_{i} =0  .
\ea
This is nothing other than energy conservation at the vertex: the energy required to create new strings balances the energy recovered from the disappearance of the old.  To see this explicitly, let us calculate the Hamiltonian of the system.  Starting first for a single $(p,q)$ string, from the Lagrangian
$S= \int d  \sigma \, {\cal L}$ defined from Eq.~(\ref{action1}), and using (\ref{F2}) and 
(\ref{sqrt}), one obtains
\ba
{\cal H}= p   F_{ \tau \sigma } + 
\dot \x  \frac{\partial {\cal L } }{\partial  \dot \x } - {\cal L} = \frac{\bm}{\la}.
\ea
Thus the tension of the $(p,q)$ string is given by $T= \bm/\la $, while the total energy of the 3-string system with junction is
\ba
E = \frac{1}{\la}\sum_i \int^{s_i(t)} d \sigma \,  \bmi  .
\ea
Eq.~(\ref{energy}) is simply $dE/dt=0$. 

Now we show that Eq.~(\ref{energy}) can be satisfied  without imposing 
extra constraints on the system, and hence that the temporal gauge can always be imposed.
Denote the spatial component of $\bar X(t)$ by $\bar \x$.
On multiplying  the spatial components of  (\ref{fcons}) by $\dot {\bar \x}$ one obtains
\ba
\label{f3}
\sum_{i} \bmi \left(  \x_{i}'  + \dot s_{i}   \dot \x_{i} \right) \cdot  \dot {\bar \x} =0 .
\ea
Now, the derivative of $\x_{i}(s_{i}(t), t)= \bar \x$ is
\ba
\label{barx}
\dot \x_{i} +\dot s_{i}   \x_{i}'  = \dot{\bar \x}  
\ea
which, combined with Eq.~(\ref{temp}), gives
\ba
\label{identity}
\x_{i}' \cdot \dot {\bar \x} = \dot s_{i}  \x_{i}'^{2} , \qquad
\dot \x_{i}\cdot \dot {\bar \x} = \dot  \x_{i}^{2}   .
\ea
Using these identities in Eq.~(\ref{f3}), one obtains
\ba
\sum_{i} \bmi     \dot s_{i}  (\x_{i}'^{2} + \dot \x_{i}^{2}) 
= \sum_{i}  \bmi    \dot s_{i}  =0  .
\ea
This indicates that Eq.~(\ref{energy}) can be satisfied automatically and 
one \emph{can} choose the temporal gauge without imposing extra constraints on the system.

Finally we note that the spatial components of (\ref{fcons}) yield the following equations for $ \bmi      (1 -\dot s_{i}^{2} )  $:
\ba
\label{seq}
\sum_{i}    \bmi      (1 -\dot s_{i}^{2} )   \x_{i}' =0  .
\ea
One immediate corollary of this equation is
\ba
\label{corollary}
\x_{1}'  \cdot ( \x_{2}' \times \x_{3}')=0  ,
\ea
which indicates that the $\x_{i}'$ are coplanar at the point of the junction.



\section{Constraints on $(p,q)$-string junction formation\label{constraints}}

In this section we apply the above formalism to study the collision of strings with charges $(p_1,q_1)$
and $(p_2,q_2)$ meeting at an angle $\alpha$ and traveling with equal and opposite velocities $v$.  When the strings collide, they may become linked by a string with charges $(p_3,q_3)=-(p_1+p_2,q_1+q_2)$.  Here we show that strong kinematic constraints apply to this process, coming from the requirement that the length of the joining string must increase in time.  We then give specific examples in the following subsections.

In \cite{Copeland:2006eh, Copeland:2006if} the same question was addressed for the collision of NG strings with tensions $\mu_1$ and $\mu_2$ (the NG limit for $(p,q)$ strings in the following analysis
can be obtained  by setting $p=0$ and $\tmu=\mu$).    There the procedure was to write the solution of the wave equation (\ref{Xeom}) in terms of ingoing and outgoing waves;
\ba
\label{ab}
\x_{i} (t, \sigma) = \frac{1}{2} \left[ \baa_i (\sigma + t) +  \bb_i(\sigma - t) \right],
\ea
where, from the temporal gauge conditions (\ref{temp})
\ba
\baa_i'^2 =  \bb_i'^2 = 1.
\ea
The amplitude of the ingoing waves $\bb_i'$ are fixed by the initial conditions and it was shown that one can solve for $\dot{s}_i$ in terms of these.  Causality (or more specifically the requirement that $|\dot{s}_i| \leq 1$) imposes that the $\mu_i$ satisfy the triangle inequalities, namely that for all $i$
\be
\nu_i \geq 0 
\ee
where, for example,
\be
\nu_1 = \mu_2 + \mu_3 - \mu_1 
\ee
(and  cyclic  permutations).

Using the equations derived in previous sections, it is straightforward to see that the procedure detailed in \cite{Copeland:2006if} goes through in exactly the same way provided {\it a)} one makes the replacement
\be
\mu_i \rightarrow \tmu_i , \qquad  (\nu_i \rightarrow \tnu_i)
\ee
in the equations of \cite{Copeland:2006if}, where $\tmu_i$ is given in terms of $p_{i}, q_{i}$ and $g_{s}$ in Eq.~(\ref{barmu}); and {\it b)} one imposes the two charge conservation conditions (\ref{pcons}) and (\ref{qcons}).   
An immediate collorolary of this is that $(p,q)$ strings meeting at a junction automatically satisfy the triangle inequalities between the $\tmu_i$, namely $\tnu_i \geq 0$.  
%
%

Also  we note that  if $g_{s}\ll 1$  then
\ba
\label{smallg}
\bmi \simeq \frac{|q_i | }{g_ s} \left( 1+ \frac{g_s^2   p_i^2 }{2   q_i^2} \right), \qquad q_i \neq 0  .
\ea
As expected, in this case the D-strings are much heavier than F-strings, by a factor of $g_s^{-1}$.
Furthermore, the binding energy of a $(p,q)$-string is almost equal to the total tension of $p$ F-strings:
$$
\bm_{(0,q)} + \bm_{(p,0)} -\bm_{(p,q)} \simeq \bm_{(p,0)} - {\cal O} (g_s)  .
$$


Now consider two strings of tension $\tmu_1$ and $\tmu_2$ parallel to
the $xy$-plane but at angles $\pm\al$ to the $x$-axis, and
approaching each other with velocities $\pm v$ in the
$z$-direction.  Before the collision (for $t<0$) we take
 \be \bx_{1,2}(\si,t)=(-\ga^{-1}\si\cos\al,
 \mp\ga^{-1}\si\sin\al,\pm vt),\label{initial} 
 \ee
where 
$\ga^{-1}=\sqrt{1-v^2}$.  As in \cite{Copeland:2006eh, Copeland:2006if}, we may distinguish two cases: the formation of an $x$-link in which the two segments of the original strings in the positive-$x$ region are joined to one end of the third, linking string, and a $y$-link, where it is the two segments in the positive-$y$ region that are joined. We consider here the formation of an $x$-link and the signs are chosen such that $\si$ increases towards the junction.
Thus
 \ba 
 \bf{a}'_{1,2}&=&(-\ga^{-1}\cos\al,\mp\ga^{-1}\sin\al,\pm v),
 \nonumber
 \\
 \bf{b}'_{1,2}&=&(-\ga^{-1}\cos\al,\mp\ga^{-1}\sin\al,\mp v). 
 \ea
At the collision, $t=0$, we suppose that the strings bind
%
with the formation of an $x$-link. We suppose the new string is oriented at an angle $\theta$ to the $x$-axis and 
moves along the $z$-direction with the velocity $u$; therefore
\ba
\x_{3}(\sigma,t)= (\gamma_{u}^{-1} \sigma \cos\theta,  \gamma_{u}^{-1} \sigma \sin\theta, u t),
\ea
and
\ba
{\bf b_{3}'}= (\gamma_{u}^{-1} \sigma \cos\theta,  \gamma_{u}^{-1} \sigma \sin\theta, -u).
\ea

As shown in \cite{ Copeland:2006if}, one finds that 
\ba
\label{uv}
\frac{u}{v} = \frac{\tan \theta}{\tan \alpha}
\ea
whereas $u$ is determined by the following equation:
\ba
\label{ueq}
(\bm_{-}^{2} \sin^{2} \alpha)  u^{4} + \left[ \bm_{3} ^{2} (1-v^{2}) + \bm_{-}^{2} (v^{2} \cos^{2} \alpha - \sin^{2} \alpha) \right] u^{2} - \bm_{-}^{2} v^{2} \cos^{2} \alpha =0
\ea
where $\bm_{\pm}= \bm_{1} \pm \bm_{2}$. It can be checked that this algebraic equation always has one positive root for $u^{2}$; furthermore $u^2<v^2$, a result which will be used later.

The condition for junction formation is $\dot s_{3} >0$; that is the length of the joining string should increase in time.  This will only be satisfied in certain domains of the $(\alpha,v)$ plane which we now determine. One easy way to calculate $\dot s_{3}$
is as follows. Eliminating $\baa_{i}'$  in terms of $\bb_{i}'$ and $\dot {\bar \x}$ in Eq.~(\ref{barx}), from the 
spatial component of Eq.~(\ref{fcons}) one obtains
\ba
\label{bprime}
\sum_{i} \bm_{i}  (1-\dot s_{i} )  \bb_{i}'= - \bm  \dot {\bar \x}  ,
\ea
where $\bm= \bm_{1} + \bm_{2} + \bm_{3}$.
Since $\bar \x(t)= \x_{3} (s_{3}(t),t)$ then 
\ba
\label{dotx3}
\dot {\bar \x} =  (\dot s_{3}   \gamma_{u}^{-1}  \cos \theta,  \dot s_{3}   \gamma_{u}^{-1} \sin \theta, u)  .
\ea
Using this in the $y$ and $z$-components of (\ref{bprime}) gives
\ba
\label{s3eq}
\dot s_{3} = \frac{ G  \bm_{+} - \bm_{3}  }{   \bm_{+}  - G  \bm_{3}   }  ,
\ea
where 
\ba
\label{beta}
G = \frac{ \gamma^{-1} \cos \alpha   }{\gamma_{u}^{-1}   \cos \theta}
=  \sqrt{  \frac{  (1-v^{2} ) ( v^{2} \cos^{2} \alpha + u^{2} \sin^{2} \alpha )     } { v^{2} ( 1-u^{2}  ) }}  .
\ea
From (\ref{ueq}) one calculates $u^{2}$, which then can be used in (\ref{beta}) and (\ref{s3eq}) to find $\dot s_{3}$.
The result is a function of $\bm_{\pm}, \bm_{3}$ and $\alpha$. 

As mentioned before, $u^2<v^2$, which implies that $G < 1$. From the triangle inequalities the denominator of (\ref{s3eq}) is always positive.  Hence the condition for junction formation to be kinematically possible,
$\dot s_{3}>0$, is equivalent to
\ba
\label{bound1}
 G > \frac{\bm_{3} }{\bm_{+} } 
\ea
which can be rewritten as
\ba
\label{f}
f(\gm^{-1}) \equiv  A_{1}  \gm^{-4} + A_{2}   \gm^{-2}  + A_{3} <0  ,
\ea
where
\ba
\label{Ai}
A_{1} &=&  \bm_{+} ^{2} \cos^{2} \alpha \left[  \bm_{3}^{2} - \bm_{+}^{2} \sin^{2} \alpha 
 - \bm_{-}^{2}   \cos^{2} \alpha  \right], \nonumber\\
A_{2}&=& 2  \bm_{+}^{2}  \bm_{-}^{2}  \cos^{2} \alpha
-    \bm_{3}^{4} - (2  \cos^{2} \alpha -1)   \bm_{+}^{2}  \bm_{3}^{2} ,
  \nonumber\\
A_{3}&=&\mu_{3}^{4} - \bm_{+}^{2}  \bm_{-}^{2}.
\ea
Condition (\ref{f}) can then be solved straightforwardly to obtain a condition on the values of $v$ for which junction formation is possible:
\be
0 \leq v^2 < v_c^2(\alpha) ,
\ee
where the critical velocity, $v_{c}$, also depends on $\tmu_{1,2}$ and hence on the charges of the two colliding strings.
Let us denote by $v_c^{\rm max}$ the maximum value taken by $v_c$ as $\alpha$ varies from $0$ to $\pi/2$.  As was shown in \cite{Copeland:2006if}, $v_c^{\rm max} \leq 1$ only if
\be
\tmu_3^2 \geq  \bm_{+} |\bm_{-} |=  |\tmu_1^2 - \tmu_2^2| .
\label{cc}
\ee
If this condition is satisfied by the tensions of the joining strings, then there is a velocity $v_c^{\rm max}$ above which the two colliding strings will pass through each other rather than forming a junction.

 We now study some physically interesting examples, namely  
collisions between $(p,q)$-strings in the low-lying energy states, such as (0,1), (1,0), $(\pm 1, \pm 1)$,\dots.


\subsection{Example 1: collision of two strings of equal tension \label{examples}}

Initially we consider the simplest case from the point of view of the formalism, namely the collision of strings of the same tension $\tmu_1=\tmu_2$ so that $\tmu_-=0$.  Examples include colliding $(1,1)$ and $(-1,1)$ strings for example.  Note that condition (\ref{cc}) is always satisfied, so that a region of
the $(\alpha,v)$ plane is always forbidden.  
Since $\tmu_-=0$ it is straightforward to see from (\ref{uv}) and (\ref{ueq}) that $u=\theta=0$.  Thus $G=\gamma^{-1} \cos\alpha$ and, provided the triangle inequalities are satisfied, the kinematic constraint (\ref{bound1}) is that
\be
\gamma^{-1} \cos\alpha > \frac{\tmu_3}{2\tmu_1}
\label{eq}
\ee
(which one can verify is identical to condition (\ref{f})).

Now take the first string, string 1, to have given charges $(p_1,q_1)$ which we take to be positive.  String 2 has the same tension and if $g_{s} \neq 1$, then from (\ref{barmu}), its charges in general are
given by $(p_2,q_2)=(\pm p_1, \pm q_1)$.  These four cases are summarised in table 1 and figure 1.

As noted in table 1, the collision of a $(p_1,q_1)$ string with an identical $(p_1,q_1)$ string (case A1) cannot lead to the formation of a joining string with charges $-2(p_1,q_1)$: this is forbidden since from (\ref{eq}) it can only take place for $v=0=\alpha$.  Physically this is expected, since identical colliding $(p,q)$ strings
are expected to intercommute with each other in the usual sense rather than forming a junction.
A related phenomenon occurs when the collision is with an antistring $-(p_1,q_1)$ (case A2): then $\tmu_3=0$ so that the linking string has zero tension.  Again we expect the strings to undergo a  standard intercommutation, in that when the two identical strings meet they
intercommute in the usual fashion, simply exchanging partners.  Indeed notice that both A1 and A2 are related to each other in that these two initial configurations are the same, just rotated by 90$^o$, the difference being that a prospective x-link for A2 is a y-link for
A1 and vice versa. The junction forms only in cases A3 and A4 and for small $g_s \ll 1$, the largest region of the $(\alpha,v)$ plane is open to the collision of $(p_1,q_1)$ with $(p_1,-q_1)$ (case A3): that is, it is much more likely to form a light joining $p$-string than a heavy joining $q$-string. The symmetry between the
cases A3 and A4 under $p\rightarrow q$ and $g_{s} \rightarrow 1/g_{s}$ is evident from the table. Physically this is expected,  since F and D-strings are symmetric under string theory S-duality where weak coupling is replaced by the strong coupling, i.e.  $g_{s} \rightarrow 1/g_{s}$.

\begin{table}

\begin{tabular}{|l|l|l|l|l|}
\hline \hline
{\bf $(p_2,q_2)$}   & {\bf $(p_3,q_3)$ } &{\bf $\tmu_3$ }& {\bf Kinematic constraint} &{\bf Case}
\\
\hline \hline
$(p_2,q_2)=(+p_1, +q_1)$ & $(-2p_1,-2q_2)$& $ \tmu_3 = 2 \sqrt{p_1^2 + \left(\frac{q_1}{g_s}\right)^2} $&intercommutation allowed with no link & A1
\\
\hline
$(p_2,q_2)=(-p_1, -q_1)$  &$ (0,0)$ &  $ \tmu_3 = 0$ & intercommutation allowed with no link& A2
\\
\hline
$(p_2,q_2)=(+p_1, -q_1)$ & $(-2p_1,0)$  &  $ \tmu_3 = 2 p_1 $&$ \gamma^{-1} \cos \alpha \geq \frac{1}{\sqrt{1+q_1^2/(g_s^2 p_1^2)}}$
& A3
\\
\hline
$(p_2,q_2)=(-p_1, +q_1)$& $(0,-2q_1)$&  $ \tmu_3 = 2 \frac{ q_1}{g_s} $  &$ \gamma^{-1} \cos \alpha \geq \frac{1}{\sqrt{1+(p_1^2  g_s^2)/\tq_1^2}}$
& A4
\\
\hline \hline
\end{tabular}

\caption{Equal tension $\tmu_1=\tmu_2$ scattering and the formation of an $x$-link.  The charges of the 1st string are fixed and positive: $(p_1,q_1)$.  The different possible charges $(p_2,q_2)$ are given in the table, as are the charges of the joining $(p_3,q_3)$ string.  Note that the formation of a $y$-link in case A1 is equivalent (by rotation by $\pi/2$) to case A2; and similarly cases A3 and A4 are related by a rotation by $\pi/2$ (see also figure 1)}

\label{table1} 

\end{table}

\begin{figure}
\includegraphics[width=0.7\textwidth]{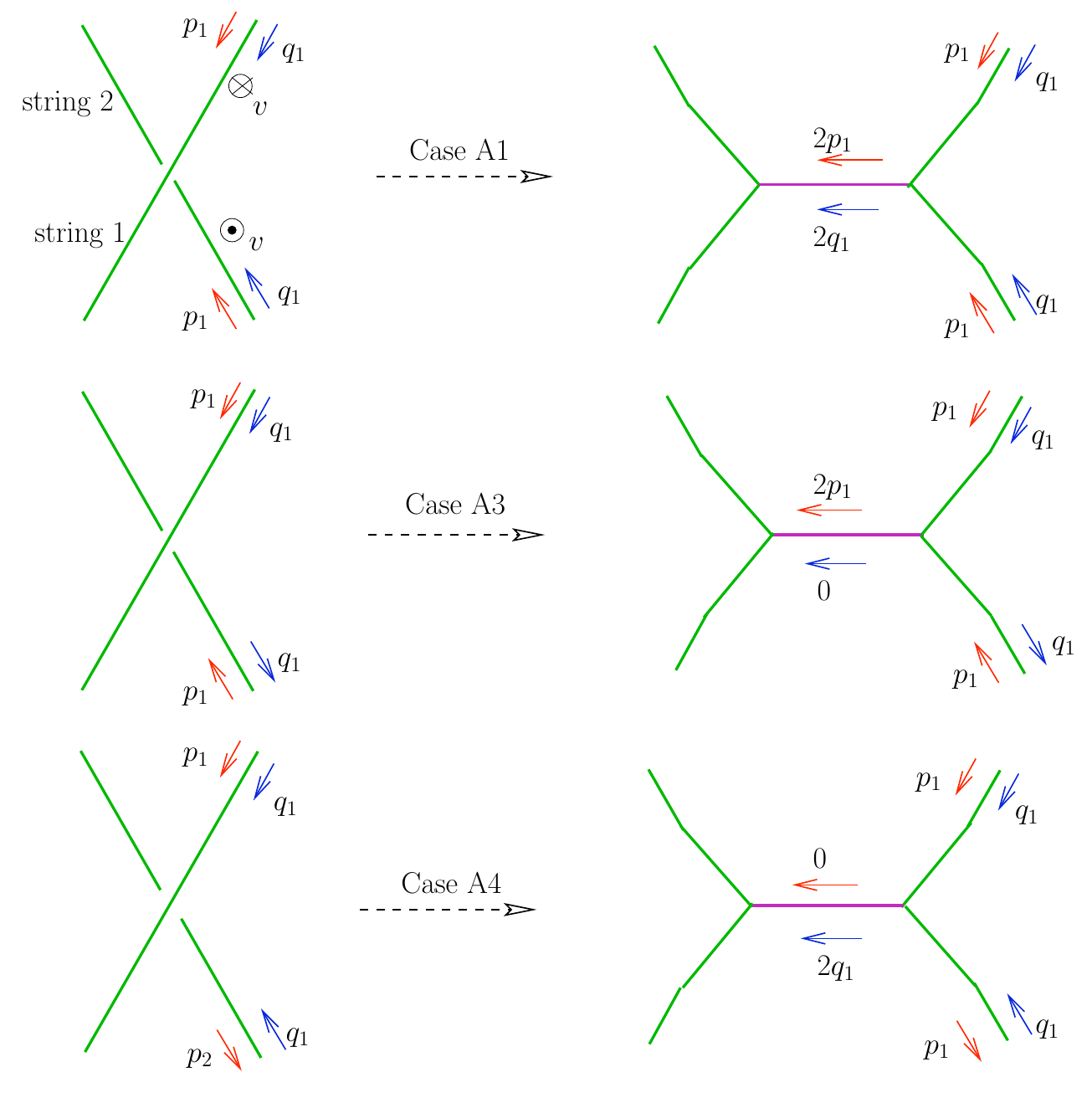}
\caption{Schematic representation of the different cases considered in the text, all with $\tmu_1=\tmu_2$ and
for $(p_1,\tq_1)$ fixed (and positive).  Note that the formation of a $y$-link in ``case A3'' is equivalent (by
rotation by $\pi/2$) to a considering the formation of an $x$-link in ``case A4''.} \label{figure1}
\end{figure}

\subsection{Example 2: collision of an F-string  with a D-string }
The collision of F, (1,0), and D (0,1)-strings is probably the most interesting example in this analysis, as it is also the basic building block for general $(p,q)$ string collisions. The third string formed at the junction is  a (1,1) string
and from Eq. (\ref{Ai}) one finds
\ba
\label{example1}
A_{1}&=& 2   g_{s}^{-3} \cos^{2}  \alpha   (2  \cos^{2} \alpha -1)    (1+g_{s} )^{2}, \nonumber\\
A_{2}&=&  2 g_{s}^{-3}  \left[  (1+g_{s}^{2}) - 2 \cos^{2} \alpha   (1+g_{s} )^{2}  \right], \nonumber\\
A_{3} &=& 4g_{s}^{-2}.
\ea

We assume that the junction is an $x$-link and $2 \cos^{2} \alpha >1$. We then see that (\ref{f}) has two
positive roots, one bigger than unity and one smaller than unity; denote the latter  by $\gm_{c}^{-2}$.
The condition (\ref{f})
for junction formation is thus equivalent to $\gm_{c}^{-2} < \gm^{-2} <1$. In terms of velocity 
this is translated into $0< v^{2} < v_{c}^{2}$ where
\ba
\label{gmc1}
v_{c}^{2} = \frac{  (1+g_{s}^{2} ) - 4 \cos^{2}  \alpha \sin^{2} \alpha (1+g_{s} )^{2}  
+ \sqrt{  (1+g_{s}^{2} )^{2}   -    4 \cos^{2}  \alpha \sin^{2} \alpha (1-g_{s}^{2} )^{2}   } }{2  \cos^{2}  \alpha  (1+g_{s} )^{2}    (2  \cos^{2} \alpha -1)     }
\ea
Note also that $v_c=0$ when  $\alpha=\pi/4$ independent of $g_{s}$. For $\alpha
> \pi/4$ no $x$-link can be formed in this case, whatever the values of
$g_s$ or $v$.

Various limits are instructive here. First consider $g_{s} \rightarrow 0$, in which case $v_{c}=1$. 
In otherwords, half the $(\alpha,v)$ plane is allowed.  Physically, this means that a very heavy D-string can always exchange partners with a light F-string.
 Furthermore, one also finds that $u\rightarrow v$ and $\theta \rightarrow \alpha$. The third string
moves with a velocity approximately equal and in a direction almost parallel to  the incoming heavy D-string. A second interesting limit is when $g_{s} \rightarrow 1$ in which case the F-string is almost as heavy as the D-string and $\bm_{-} \rightarrow 0$. This limit was studied
above since $\tmu_1=\tmu_2$ and one finds
$\gamma^{-1} \cos \alpha>1/\sqrt 2$, and $u\sim \theta \sim 0$.

\subsection{Example 3: collision of an F-string  with a $(-1,1)$ string}

A further interesting example is the collision of  (1,0) and $(-1,1)$ strings, with the third string being a
D-string. In this case
\be
\tmu_1=1, \qquad \tmu_2=\sqrt{1+\frac{1}{g_s^2}}, \qquad \tmu_3=\frac{1}{g_s} \qquad \Rightarrow \qquad
\tmu_3^2 = |\tmu_1^2 - \tmu_2^2|
\ee
so that the condition in (\ref{cc}) is saturated.  As discussed in \cite{Copeland:2006if} there is no bound on $v$ and one can show that 
\be
\label{Bound1}
\sin^2 \alpha < \frac{1}{2}\left( 
1 + \frac{g_s}{ \sqrt{1+g_s^{2} }}    \right)   
 \ee


The mirror image of the above example  is the collision of (0,1) and $(1,-1)$ strings, where the
third string is a light F-string.  Again this case satisfies the bound (\ref{cc}) so that there is no constraint on $v$.  This time, however,
\be
\label{Bound2}
\sin^2 \alpha < \frac{1}{2}\left( 
1  + \frac{1}{ \sqrt{1+g_s^{2} }}    \right)   
 \ee
 The symmetry between (\ref{Bound1}) and (\ref{Bound2}) under $g_{s} \rightarrow 1/g_{s}$ is evident.
 As explained before, this is a consequence of symmetry between F and D-strings under S-duality 
 where $g_{s} \rightarrow 1/g_{s}$.
 Now a much larger region of the $(\alpha,v)$ plane is therefore open when the joining string is lighter.

\section{Collisions in a warped background\label{warped}}

Our analysis for the collision of cosmic strings in a flat space-time can also be generalized to the collision of cosmic superstrings in a warped throat. This is of great interest because in models of warped brane inflation \cite{Kachru:2003sx}
the inflation takes place inside a  warped throat and cosmic superstrings produced at the end of inflation are located at the bottom of the throat.

To be specific, we study the collision of $(p,q)$-strings in the Klebanov-Strassler  (KS) throat 
\cite{Klebanov:2000hb} which is a warped deformed conifold.
At the tip of the throat, where we assume the strings are located, the internal geometry ends on a round three-sphere and the metric is given by
\ba
\label{metric0}
ds^2 =  h^2 \eta_{\mu \nu} dx^{\mu} dx^{\nu}+
  g_s M \alpha'(d\psi^2 +\sin^2 \psi\, d \Omega_2 ^2),
\ea
where $h$ is the warp factor at the bottom of the throat. Here $\psi$ is the usual polar coordinate on a $S^{3}$, ranging from 0 to $\pi$, and $M$ is the number of Ramond-Ramond $F_{(3)}$
fluxes turned on inside this
$S^{3}$. At the tip of the throat the two-form $C_{(2)}$ corresponding to the three form $F_{(3)}$ is given by
\ba
\label{C2}
C_{(2)}=  M \alpha'   \left(\psi-\frac{\sin (2\psi)}{2} \right) \sin \theta\, d \theta \, d\phi  .
\ea

As studied in \cite{Firouzjahi:2006vp}, (see also \cite{Thomas:2006ud, Firouzjahi:2006xa} for its dual prescription), a $(p,q)$-string in the KS  throat is constructed from a wrapped D3-brane 
with $q$ units of magnetic fluxes
and $p$ units of electric fluxes turned on in its world volume. The D3-brane is wrapped  around a $S^{2}$
inside the $S^{3}$, where each $S^{2}$ is determined by a slice of constant $\psi$, given by
\ba
\label{psi}
\psi= \frac{\pi  p}{M}  .
\ea
After integrating the contribution from the 
wrapped dimensions of the brane, and imposing the conformal gauge, 
the action for a $(p,q)$-string extended along the $z$-direction is:
\ba
\label{action2}
S= \int dt\, dz\, \left(-\Delta \sqrt{ h^4   \x'^{2} (1-\dot \x^{2})    - \la^2 F_{0z}^2} +
\Omega F_{01} 
\right)  ,
\ea
where, as opposed to the flat metric, there is now a non-zero contribution from the Chern-
Simons term;
\ba
\label{Omega}
\Omega \equiv \la \mu_3 \int_{S^2} C_{(2)} = \frac{M}{\pi} \left[ \frac{\pi  p}{M} - 
\frac{1}{2} \sin \left(  \frac{2 \pi  p}{M}\right)   \right]
\ea
and
\ba
\label{delta}
\Delta &\equiv& \mu_3 g_s^{-1} \int_{S^2} \sqrt{g_{\theta \theta} g_{\phi \phi} + \la^2 F_{\theta \phi}^2 }\nonumber\\
&=&\la^{-1} \sqrt{\frac{ M^2}{\pi^2} \sin^4 \left( \frac{\pi  p}{M}  \right) +\frac{q^2}{g_s^2} }.
\ea
Here $g_{\theta \theta}$ and $g_{\phi \phi}$ are the angular parts of the metric of $S^{2}$ in 
(\ref{metric0}) and $\mu_{3}= 1/(2 \pi)^{3} \alpha'^{2}$ is the D3-brane charge.  Note that if we
turn off the Chern-Simons term, $\Delta$ reduces to the tension of q coincident D-strings given
by $|q|/\lambda g_{s}$.

Comparing the action in Eq.~(\ref{action2}) with the action of a string in a flat background,  
$\Delta$ plays the role of the bare tension, i.e. $\mu_{i}$ as in Eq.~(\ref{action}). Furthermore, the term containing $\Omega$ in Eq.~(\ref{action2}) corresponds
to a modification to the conjugate electric charge, $p =\delta {\cal L}/ \delta F_{ t z}$ , due to the background flux.  Constructing the Hamiltonian, one can check that the tension of the $(p,q)$-
strings is equal to $\bm/\lambda$ where
\ba
\label{E}
\bm= {h^2} \sqrt{ \frac{q^2}{g_s^2} + \frac{ M^2}{\pi^2}
 \sin^2 \left(  \frac{\pi p}{M}\right) }.
\ea
In the limit where $M\rightarrow \infty$, the formula above reduces to the tension of a $(p,q)$
strings in a flat background given by Eq.~(\ref{barmu}). This is expected, since in this limit the size
of $S^{3}$ at the bottom of the throat is very large and one effectively deals with a flat background.
Furthermore, the tension of string is warped by two powers of warp factor, as expected.

The formation of a junction in a KS throat at the static level was studied in \cite{Dasgupta:2007ds}, where
it is shown that the method of wrapped D3-brane can be used for  the collision of $(0,q_{1})$ and $(p_{2}, q_{2})$ strings.
Here we would like to study the collision of an F-string and a D-string in the KS throat, which would be the generalization of F and D-strings collision studied in previous section. 

Following the same strategy as in the flat background, the action of (0,1), (1,0) and $(-1,-1)$ strings
forming a junction is 
\ba
\label{action3}
S&=& -\Delta_{i}    \int  d  \tau  \, d \sigma \, \left( \sqrt{ h^4   \x'^{2} (1-\dot \x^{2})
- \la^2 { F_{\tau\sigma}^i} ^2}  + \Omega_{i}   F_{\tau\sigma}^i     \right)    
\theta \left(s_{i}(\tau)-\sigma \right)   \nonumber\\
&+& \sum_{i}  \int d  \tau  \left(    f_i . \left[  X_i(s_i (\tau),  \tau )-\bar X(\tau)  \right]  +
g_i  \left[  A^i_\tau (s_i(\tau),\tau)+\dot s_i   A^i_\sigma(s_i(\tau),\tau)
 - \bar A(\tau) \right] \right)  .
\ea
The results in sections \ref{junctions} and \ref{temporal} formally go through with $\bm_{i}$ now given in Eq.~(\ref{E}) and
\ba \label{barmuwarped}
\bm_{1} ={h^{2}  }  g_{s}^{-1} , \quad
\bm_{2}= {h^{2}  }   \frac{M}{\pi}  \sin  \frac{ \pi }{M}, \quad
\bm_{3}=  {h^{2}  }    \sqrt{ g_{s}^{-2} + \frac{ M^2}{\pi^2}
 \sin^2   \frac{\pi }{M}}.
\ea

Constructing the coefficients $A_{1}, A_{2}$ and $A_{3}$ from Eq.~(\ref{Ai}) 
one can check that $A_{i}$ have the same value as in Eq.~(\ref{example1})
if  $g_{s} \rightarrow \bar g_{s}$,  where
\ba
\label{barg}
\bar g_{s} = \frac{M}{\pi}  \sin \left(  \frac{ \pi }{M}\right)    g_{s}  .
\ea
This indicates that the bound on the incoming velocity, $v_{c}$, is given by the same expression
as in Eq.~(\ref{gmc1}) where now $g_{s}$ is replaced by $\bar g_{s}$.  

\section{Conclusions\label{conc}}

\noindent The very attractive brane inflation scenario leads naturally to the formation of cosmic $(p,q)$-superstrings.  In this paper we have studied the behaviour of these strings at a three-string junction, and the constraints that determine whether colliding strings can exchange partners.  Constraints very similar to those found earlier for NG strings apply here too, but in addition to the energy-momentum conservation constraint (\ref{fcons}) we also have two other conservation laws, (\ref{pcons}) and (\ref{qcons}).  Moreover the parameter $\tmu_i$ appearing in (\ref{fcons}) is the tension of the $(p_i,q_i)$-string rather than $\mu_i$, the tension of $q_i$ coincident D-branes which multiplies the action in (\ref{action}).  
%
%
%
We also briefly examined the effect of non-trivial background geometry, in particular the effect of warping in a KS throat.  Here we showed that the kinematic constraints are again of the same form but with transformed effective-tension parameters, given by (\ref{barmuwarped}).

\begin{figure}
\includegraphics[width=0.4\textwidth]{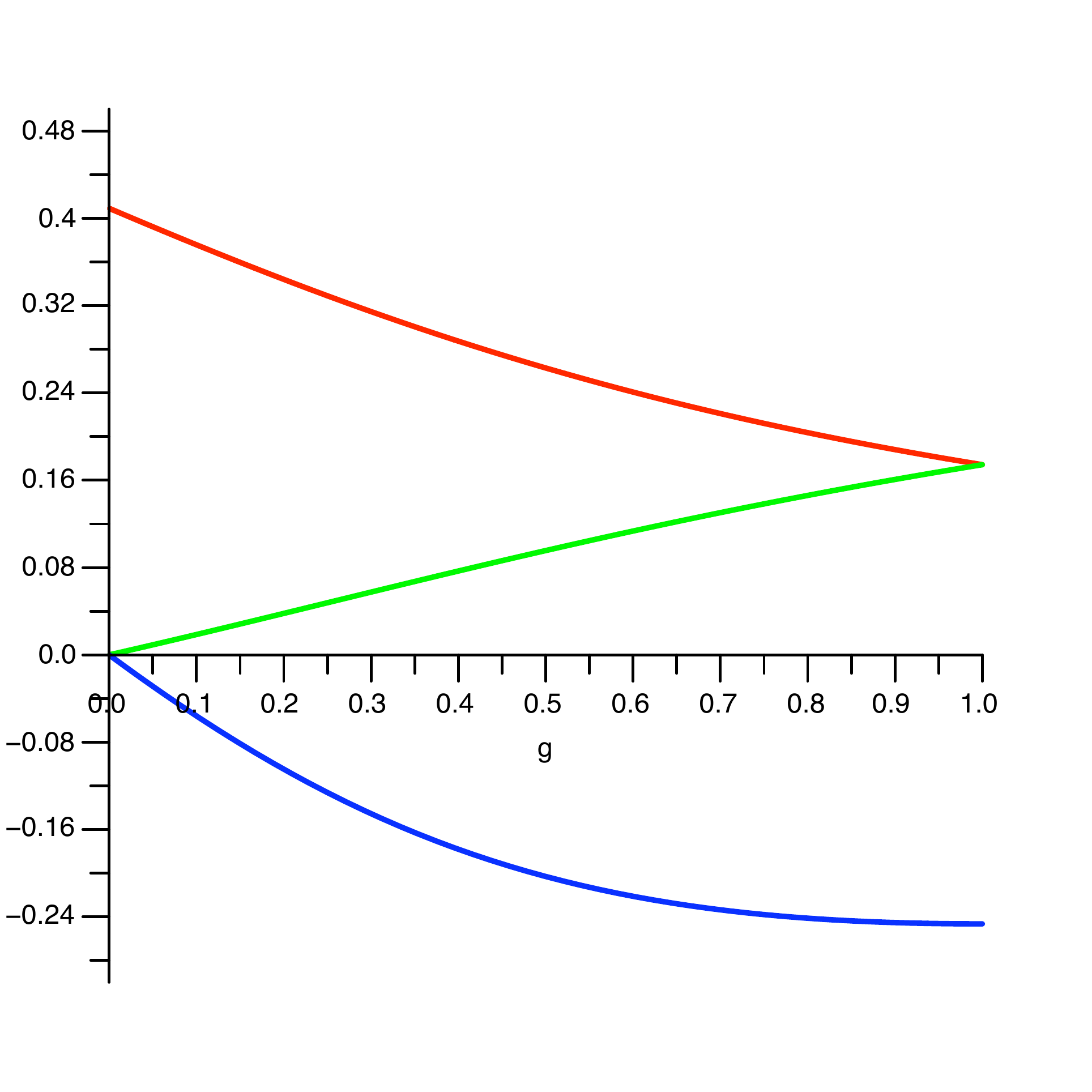}
\caption{$\langle \dot{s}_i \rangle$ as a function of $g_{s}$.  Red: $i=1$.  Blue $i=3$, green $i=2$.} \label{fig:sdot}
\end{figure}

Our results may be important in studies of the evolution of a network of $(p,q)$-strings \cite{Tye:2005fn}.   For instance, one can investigate some average properties of a network, such as $\langle \dot{s}_i \rangle$ and $\langle \dot{\bx}_i^2 \rangle$ as discussed in \cite{Copeland:2006if} for NG strings.  It is easy to see that the explicit expressions given in that paper for, for example $\langle \dot{s}_i \rangle$,  go through to $(p,q)$ junctions provided one replaces $\mu_i \rightarrow \tmu_i$.  Therefore we again expect vertices move along the strings in such a way as to increase the length of the lightest strings: This is shown in figure
\ref{fig:sdot} where we plot $\langle \dot{s}_i \rangle$ for
\be
 (p_1,q_1)=(+1,0)  , \qquad  (p_2,q_2)=(0,\pm 1) \qquad  (p_3,q_3)=(-1,\mp 1) \ee
as a function of $g_s$.   (Note that these are chosen so that  $\tmu_1 \leq \tmu_2 < \tmu_3
$ with the equality holding when $g_s=1$). 

There are a number of directions in which this work could be taken. They include using our new results to look at the effect of lensing on string junctions, and to look for a class of exact loop configurations that would allow us to determine the distinct gravitational wave emission from such objects. Perhaps the most significant use though would be as an input into detailed simulations of a network of $(p,q)$ strings, because as we have seen they throw up novel features that are not present in the case of ordinary abelian cosmic strings. 

\begin{acknowledgments}
{We would like to thank K. Dasgupta for useful discussions.
The work of H.F. is supported by NSERC.}
\end{acknowledgments}


\end{document}